\documentclass{article}

\usepackage{arxiv}

\usepackage[utf8]{inputenc} 
\usepackage[T1]{fontenc}    
\usepackage[hidelinks]{hyperref}       
\usepackage{url}            
\usepackage{booktabs}       
\usepackage{amsfonts}       
\usepackage{nicefrac}       
\usepackage{microtype}      
\usepackage{cleveref}       
\usepackage{graphicx}
\usepackage{doi}
\usepackage{longtable}
\usepackage{supertabular}
\usepackage[english]{babel}
\usepackage[table,xcdraw]{xcolor}
\usepackage{breakurl}

\graphicspath{ {./} }
\begin{document}

\title{Implementation of communication media around a mixed reality experience with HoloLens headset, as part of a digitalization of a nutrition workshop}
\renewcommand{\shorttitle}{Communication media around a mixed reality experience}

\author{
\href{https://orcid.org/0000-0002-9298-7492}{\includegraphics[scale=0.06]{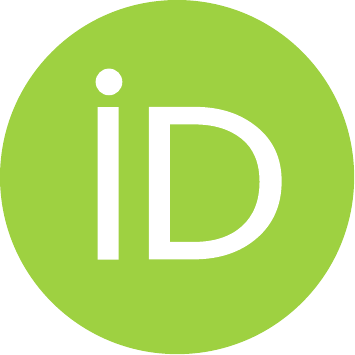}
    \hspace{1mm}Owen~K.~Appadoo}\thanks{Université Clermont Auvergne, IUT Clermont Auvergne site du Puy-en-Velay, Institut Pascal, 8 Rue Jean-Baptiste Fabre CS 10219 43009 Le Puy-en-Velay}\\
    Université Clermont Auvergne,\\
    Clermont Auvergne INP,\\
    CNRS, Institut Pascal\\
    F-63000 Clermont-Ferrand, France\\
    \texttt{owen.appadoo@uca.fr}
    \And
    \href{https://orcid.org/0000-0002-5264-2027}{\includegraphics[scale=0.06]{./orcid.pdf}
    \hspace{1mm}Hugo~Rositi}\thanks{corresponding author}\\
    Université Clermont Auvergne,\\
    Clermont Auvergne INP,\\
    CNRS, Institut Pascal\\
    F-63000 Clermont-Ferrand, France\\
    \texttt{hugo.rositi@uca.fr}
    \And 
    Sylvie Valarier\\
    Services de Chirurgie Bariatrique/Nutrition,\\
    C.H. Emile Roux du Puy-en-Velay,\\
    12 boulevard Docteur Chantemesse,\\
	F-43000 Le Puy-en-Velay, France\\
	\texttt{sylvie.valarier@ch-lepuy.fr}
    \And 
    Marie-Claire Ombret\\
    Unité transversale d'éducation du patient,\\
    C.H. Emile Roux du Puy-en-Velay,\\
    12 boulevard Docteur Chantemesse,\\
	F-43000 Le Puy-en-Velay, France\\
	\texttt{marieclaire.ombret@ch-lepuy.fr}
    \And 
    \'Emilie Gad\'ea \\
    Unité de Recherche Clinique,\\
    C.H. Emile Roux du Puy-en-Velay,\\
    12 boulevard Docteur Chantemesse,\\
	F-43000 Le Puy-en-Velay, France\\
	\texttt{responsable.rechercheclinique@ch-lepuy.fr}
    \And 
    Christine Barret-Grimault\\
    Services de Chirurgie Bariatrique/Nutrition,\\
    C.H. Emile Roux du Puy-en-Velay,\\
    12 boulevard Docteur Chantemesse,\\
	F-43000 Le Puy-en-Velay, France\\
	\texttt{christine.barret@ch-lepuy.fr}
    \And
    \href{https://orcid.org/0000-0001-5352-8237}{\includegraphics[scale=0.06]{./orcid.pdf}
    \hspace{1mm}Christophe Lohou}\\
    Université Clermont Auvergne,\\
    Clermont Auvergne INP,\\
    CNRS, Institut Pascal\\
    F-63000 Clermont-Ferrand, France\\
	\texttt{christophe.lohou@uca.fr}
}

\renewcommand{\headeright}{Appadoo \textit{et~al.}, 2023.}
\renewcommand{\undertitle}{Multimedia communication article}

\hypersetup{
    pdftitle={Communication article},
    pdfsubject={multimedia},
    pdfauthor={Hugo Rositi},
    colorlinks=true,
    linkcolor=black,          
    citecolor=black,        
    filecolor=black,         
    urlcolor=black        
}
\date{\today}
\maketitle
\begin{abstract}
	The release of Microsoft's HoloLens headset addresses new types of issues that would have been difficult to design without such a hardware. 
	This semi-transparent visor headset allows the user who wears it to view the projection of 3D virtual objects placed in its real environment. 
	The user can also interact with these 3D objects, which can interact with each other. The framework of this new technology is called mixed reality. 
	We had the opportunity to numerically transform a conventional human nutrition workshop for patients waiting for bariatric surgery by developing  
	a software called \textbf{HOLO\_NUTRI} using the HoloLens headset. Unlike our experience of user and conventional programmer specialized in the development 
	of interactive 3D graphics applications, we realized that such a mixed reality experience required specific programming concepts quite different from 
	those of conventional software or those of virtual reality applications, but above all required a thorough reflection about communication for users. 
	In this article, we propose to explain our design of communication (graphic supports, tutorials of use of material, explanatory videos), a step which
	 was crucial for the good progress of our project. The software was used by thirty patients from Le Puy-en-Velay Hospital during 10 sessions of one 
	 hour and a half during which patients had to take in hand the headset and software \textbf{HOLO\_NUTRI}. We also proposed a series of questions to patients to 
	 have an assessment of both the adequacy and the importance of this communication approach for such experience. As the mixed reality technology is 
	 very recent but the number of applications based on it significantly increases, the reflection on the implementation of the elements of 
	 communication described in this article (videos, exercise of learning for the use of the headset, communication leaflet, \textit{etc.}) can help developers 
	 of such applications.
\end{abstract}


\section{Context}
\label{secContext}
\subsection{Introduction}
\label{subsecIntroduction}
One of the patient therapeutic education workshops of the Hospital Center Emile Roux nutrition department in Le Puy-en-Velay, France, is to make patients in need of bariatric surgery
(reducing the stomach volume) aware of the changes in their diet after surgery. Patients have to compose a standard menu after surgery using printed cards on a table (food/associated 
quantity), then the nutrition team analyzes their choices and delivers qualitative and quantitative messages. We call this workshop \textbf{CONV\_WORKSHOP}.\\ 

A team of computer scientists collaborated with the nutrition team to digitally transform their menu composition workshop by using the HoloLens headset, recently 
introduced by Microsoft. We call \textbf{DIG\_WORKSHOP} this digital workshop version. These scientists then developed a computer application, called \textbf{HOLO\_NUTRI}, 
to allow patients to compose their menu while using the headset. Synthetic 3D elements (such as a cube) are superimposed on the semi-transparent visor without occulting the vision through 
the headset. Unlike a direct approach to augmented reality, the user can interact with these synthetic objects, they can also interact with each other and with the environment, 
this new technological framework is called \textit{mixed reality}.

\par In addition to the computer design for the development of \textbf{HOLO\_NUTRI} operating various software and libraries, a specific development is required to provide the most
appropriate user experience, for example, an arrangement of manipulable 3D objects at a distance from the user (aspects developed in the article \cite{Rositi2020}). The obligation 
to propose an appropriate communication for a good comprehension of both the material and the software \textbf{HOLO\_NUTRI} quickly appeared to us, so that the patients could use 
this software and understand the exploitation of the headset (superimposition of 3D objects, gestures to perform, \textit{etc.}) in less than 1h30. 

In this article, we describe how we digitalized such a workshop using both holographic headsets and our produced software \textbf{HOLO\_NUTRI}. More specifically, we justify all the 
steps of communication media necessary for the success of this project. These communication elements are offered in three different forms: paper support (leaflet), multimedia 
support (video sequences), software support (learning exercise). More precisely, four leaflets, four video sequences and a "learning exercise" software, named \textbf{LEARN\_EX}, have been 
developed for this purpose. We think that this experiment could be adopted for any communication using this technology, in our case it represents
about a quarter of the overall preparation time of the project, the remaining concerns the encoding of our software, so it is an important element to take into account.

\par The article is structured in the following way: in section~\ref{subsecProgressConv}, we first describe the progress of the conventional \textbf{CONV\_WORKSHOP} workshop in order to understand later
the complexity of its digitization exploiting the mixed reality. Then we explain the objectives to be proposed in the context of this digitization (section~\ref{subsubsecWorkshopDig}), and justify the 
choice of mixed reality (section~\ref{subsubsecMixed Reality}) to achieve them in an optimal way. We briefly describe the features of the \textbf{HOLO\_NUTRI} application 
(section~\ref{subsubsecHolosoft}), and the correct way to dispose the headset (section~\ref{subsubsecsetDig}). Then, we detail the steps of the numerical workshop \textbf{DIG\_WORKSHOP} 
(section~\ref{secDIGWORKSHOP}), in order to 
understand the relevance of the various elements of communication. Section~\ref{secPrelimDIGWORKSHOP} presents the global communication process for setting up the workshop \textbf{DIG\_WORKSHOP}; this same
process is more detailed in section~\ref{secDescglobalcomm}. In this same section, we will focus on the development of video sequences for the presentation of the headset, the actions to be performed 
and associated gestures (section~\ref{subsecSteps1a4}), the presentation sequence of the \textbf{LEARN\_EX} learning exercise (section~\ref{subsecLEARNEX}), and finally the design of an information leaflet used 
in \textbf{HOLO\_NUTRI} (section~\ref{subsecNutri}). Section~\ref{secResults} will summarize the assessments made by the thirty patients concerning the communication. We conclude in section~\ref{secConclusion}.

\subsection{Progress of the conventional nutrition workshop}
\label{subsecProgressConv}
\begin{figure}[!t]	
	\centering
	\includegraphics[width=0.5\textwidth]{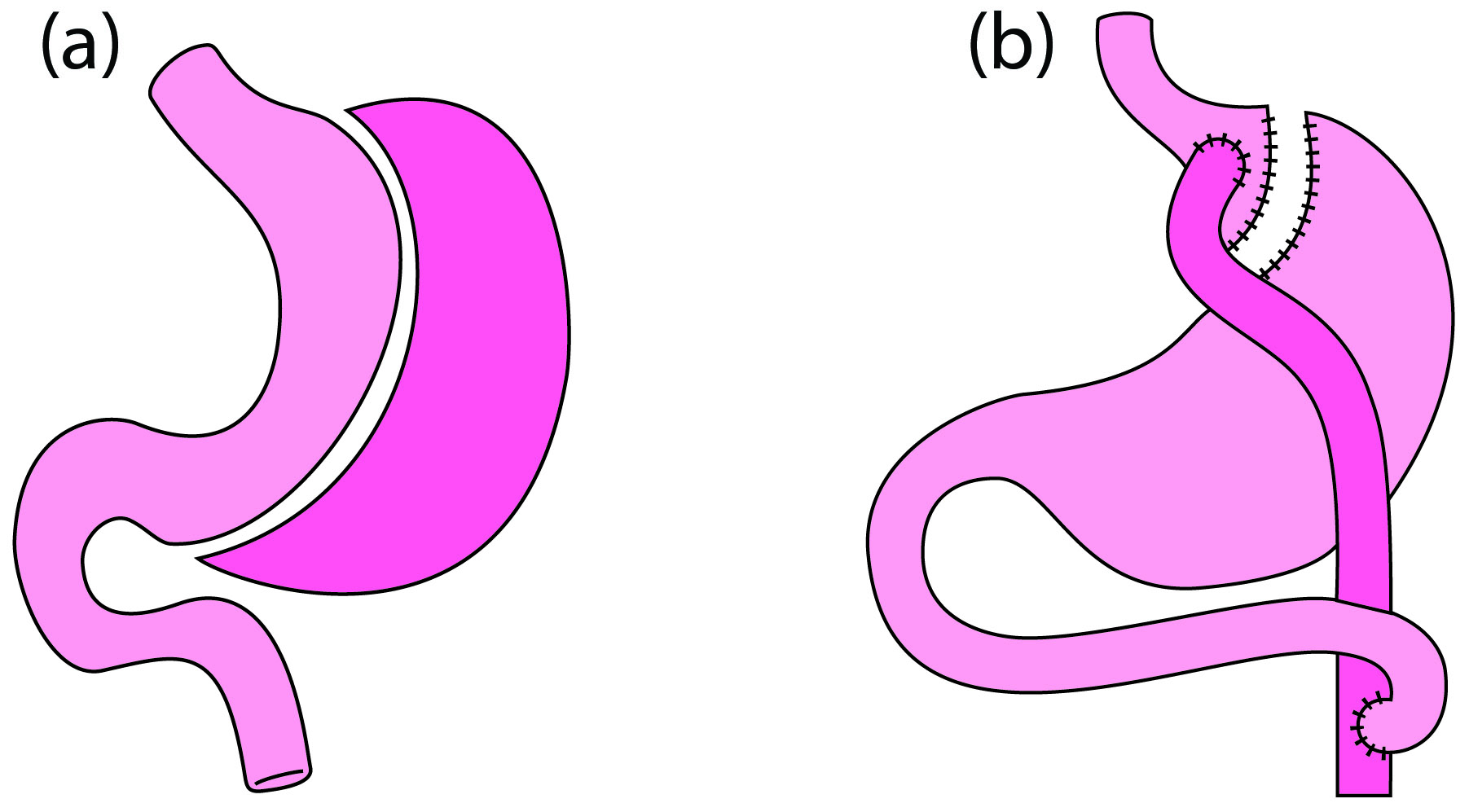}
	\caption{Both types of interventions considered : (a) Sleeve Surgery – the dark part is removed, (b) Bypass Surgery – the dark part represents the middle part of the small intestine 
	which is a shortcut for food flow. Figure from \cite{Rositi2020}.}
	\label{figure1}
\end{figure}

The nutrition team composed of endocrinologists, digestive surgeon, nurse coordinator psychologist and dietician receives patients who will undergo a 
bariatric surgery. This surgery aims at a weight loss of the patient. For that, two surgical solutions are considered: either the reduction of the volume of 
the stomach "Sleeve Surgery" (Figure~\ref{figure1}(a)), or to shunt the normal way of progression of the foods "Bypass Surgery" (Figure~\ref{figure1}(b)).  
The proposed workshop \textbf{CONV\_WORKSHOP} aims to make the patients understand (on average 6 by group) the difference of diet before and after surgical 
intervention. The workshop lasts one hour and half.

\par The workshop is organized according to the 9 steps described below (summarized in Table 1):
Introduction of the session (step 1); debate around the question «Feeding ourselves, eating, when we have been 
under a bariatric surgery, what does it bring to mind ?» (step 2); initiation of the workshop to discuss with patients about 
their daily diet to adopt after intervention (step 3); patients are then invited to compose their menu by choosing cards representing different foods according different quantities, 
the images come from the SU.VI.MAX study \cite{Hercberg2004,Hercberg2004b} (step 4); collection by the clinical team of the menus developed by patients (step 5); collegial discussion
 of the choice made by each patient (step 6); the dietician then gives each patient some advices based on the foods and quantities she or he has chosen (step 7); dissemination of 
 quantitative and qualitative messages (step 8). The qualitative messages relate on the need to vary the number of dishes, the duration of the meal, the chewing time; quantitative 
 messages are about reducing the amount of food compared to a meal before surgery. Finally, patients complete forms to evaluate the session (step 9).

\begin{table}
	\centering
	\caption{Description of both steps and actions during conventional workshop \textbf{CONV\_WORKSHOP}. Actors of all actions are identified by specific colors detailed at the end of this table.}
	\label{table1}
	\begin{tabular}{|l|l|} 
	\hline
	\multicolumn{2}{|l|}{1. Session introduction} \\
	\hline
	& \cellcolor[HTML]{4f81bd}{\color[HTML]{ffffff}{1.1. Presentation of the protocol and signature of the consent forms}} \\ 
	\hline
	& \cellcolor[HTML]{9bbb59}1.2. Signature by patient \\
	\hline
	& \cellcolor[HTML]{fabf8f}1.3. Group presentation (clinicians / patients) \\
	\hline
	& \cellcolor[HTML]{4f81bd}{\color[HTML]{ffffff}{1.4. Presentation of the objectives of the session}} \\
	\hline
	\multicolumn{2}{|l|}{}  \\ 
	\hline
	\multicolumn{2}{|l|}{\cellcolor[HTML]{fabf8f}\begin{tabular}[c]{@{}l@{}}2. Debate around the question «Feeding ourselves, eating, when we have been \\ under a bariatric surgery, what does it bring to mind?»\end{tabular}}  \\ 
	\hline
	\multicolumn{2}{|l|}{}  \\ 
	\hline
	\multicolumn{2}{|l|}{\cellcolor[HTML]{4f81bd}{\color[HTML]{ffffff}\begin{tabular}[c]{@{}l@{}}3. Beginning of the workshop: Initial question "after the surgery, how  \\ to eat daily? "\end{tabular}}} \\ 
	\hline
	\multicolumn{2}{|l|}{}  \\ 
	\hline
	\multicolumn{2}{|l|}{4. Carrying out the workshop experience with cards}  \\ 
	\hline
	& \cellcolor[HTML]{4f81bd}{\color[HTML]{ffffff}{4.1. Presentation of the workshop with cards}} \\ 
	\hline
	& \cellcolor[HTML]{9bbb59}4.2. Food choices via cards \\
	\hline
	& \cellcolor[HTML]{4f81bd}{\color[HTML]{ffffff}{4.3. Register quantities in a table}} \\
	\hline
	\multicolumn{2}{|l|}{}  \\ 
	\hline
	\multicolumn{2}{|l|}{\cellcolor[HTML]{fabf8f}\begin{tabular}[c]{@{}l@{}}5. Discussion around the choices of each patient with the whole group \\ (collecting information of each patient)\end{tabular}}  \\ 
	\hline
	\multicolumn{2}{|l|}{}  \\ 
	\hline
	\multicolumn{2}{|l|}{\cellcolor[HTML]{4f81bd}{\color[HTML]{ffffff}\begin{tabular}[c]{@{}l@{}}6. Highlighting important messages (in relation to the choices made  \\  by each patient) \end{tabular}}}  \\ 
	\hline
	\multicolumn{2}{|l|}{}  \\ 
	\hline
	\multicolumn{2}{|l|}{\cellcolor[HTML]{fabf8f}\begin{tabular}[c]{@{}l@{}}7. Validation of exact knowledge / readjustment of erroneous knowledge  \\ (advices to each patient)\end{tabular}}  \\ 
	\hline
	\multicolumn{2}{|l|}{}  \\ 
	\hline
	\multicolumn{2}{|l|}{\cellcolor[HTML]{4f81bd}{\color[HTML]{FFFFFF}\begin{tabular}[c]{@{}l@{}}8. Informative messages (summarized by the workshop host)\end{tabular}}}  \\ 
	\hline
	\multicolumn{2}{|l|}{}  \\ 
	\hline
	\multicolumn{2}{|l|}{\cellcolor[HTML]{9bbb59}\begin{tabular}[c]{@{}l@{}}9. Forms (assessment knowledge + satisfaction)\end{tabular}}  \\ 
	\hline
	\end{tabular}
	\begin{tabular}{llllll}
		\cellcolor[HTML]{4f81bd} & \textit{Clinical team action} & \cellcolor[HTML]{fabf8f} & \textit{Clinical team + patients action} & \cellcolor[HTML]{9bbb59} & \textit{Patients action} \\
		\end{tabular}
\end{table}

\par We stress on the importance of oral communication throughout this workshop. The gestural interaction of the patients, on the other hand, is done only when they take the cards 
representing the food. 

\subsection{Our incentive : a digitalization of the workshop by mixed reality}
\label{subsecIncentive}
\leftskip=0.6cm
\subsubsection{Workshop digitalization}
\label{subsubsecWorkshopDig}
\leftskip=0cm

We have proposed a software version (digitalization) of the conventionnal workshop \textbf{CONV\_WORKSHOP}, previously described in section~\ref{subsecProgressConv}. This new workshop 
\textbf{DIG\_WORKSHOP} aims to propose:
\begin{itemize}
	\item the digital version of menu composition using a computer application \\(\textbf{HOLO\_NUTRI}),
	\item a more detailed analysis of the developed menu (several quantities associated with food, simulation of the meal time, taking into account the chewing time, \textit{etc.}),
	\item a similar approach, in the computer application, of the patient's gesture when she or he grabs the card representing the food and the associated quantity in 
	\textbf{CONV\_WORKSHOP}.
\end{itemize}
The first two items above can be done using a conventional computer and software development. On the other hand, we have sought the most appropriate solution as regards to the 
implementation of the gesture to catch the numerical model corresponding to the printed card of the food, or even to reinforce this learning. To do this, we then opted for a 
solution setting up a mixed reality framework that we define below.

\leftskip=0.6cm
\subsubsection{Mixed Reality}
\label{subsubsecMixed Reality}
\leftskip=0cm

\begin{figure}[!t]	
	\centering
	\includegraphics[width=0.5\textwidth]{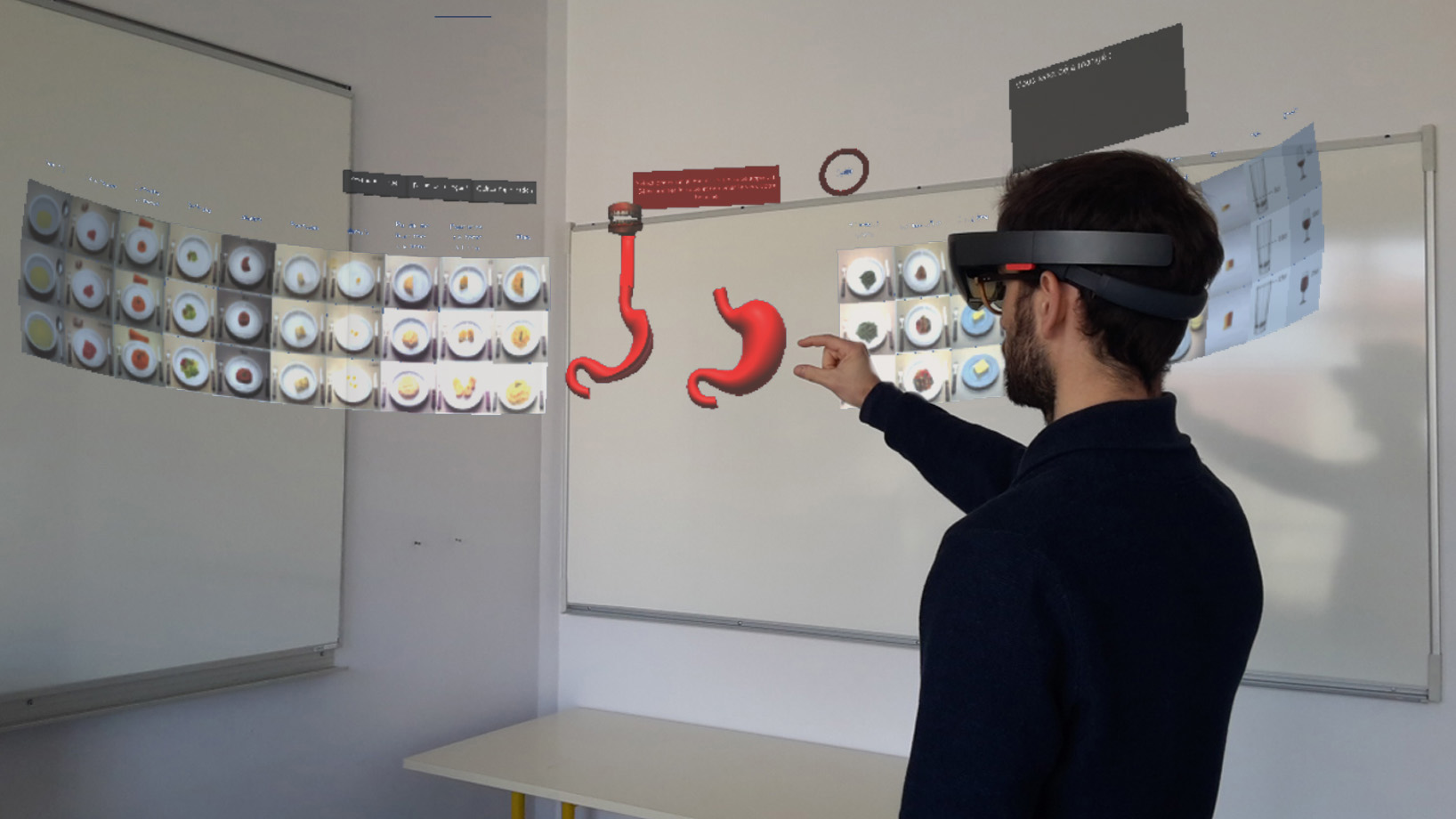}
	\caption{Photomontage illustrating the visualization of the headset wearer in his environment.}
	\label{figure2}
\end{figure}

Microsoft has recently released the HoloLens headset \cite{hololens2020}. In addition to scanning the real environment around the user, this headset allows to display virtual 
3D objects (called holograms) superimposed on the visualization of the real environment, and to interact with these same objects thanks to different sensors headset 
(Figure~\ref{figure2}).

Therefore, it is the physical aspect (collisions, \textit{etc.}) which is put forward for the exploitation of the mixed reality. The user can interact with his environment using the 
GGV triplet (Gaze, Gesture, Voice) \cite{GGV2020}, by triggering actions in the computer application. Indeed, the integrated gyroscope computes the movements of the head - Gaze, it 
should be noted that it is the orientation of the head which is taken into account and defines this notion of gaze, the eyes are not tracked; the depth cameras of the headset detects 
certain gestures of the user (hand fingers tight, open hand, \textit{etc.}) - Gesture; a microphone can hear the words or sounds of the user - Voice.

\par More and more applications using HoloLens within an industrial framework (automobile, architecture, engineering, medical) are proposed, to give, for example, information on a 
product during industrial maintenance or bring anatomical information during surgical operations \cite{Lohou2019,Lohou2019b}.
In the next section, we describe our mixed reality software.

\subsection{Our solution}
\label{subsecOursolution}
\leftskip=0.6cm
\subsubsection{Description of \textbf{HOLO\_NUTRI} software}
\label{subsubsecHolosoft}
\leftskip=0cm

The \textbf{HOLO\_NUTRI} software is composed of two stages. 

\par
\begin{figure}[!t]	
	\centering
	\includegraphics[width=0.5\textwidth]{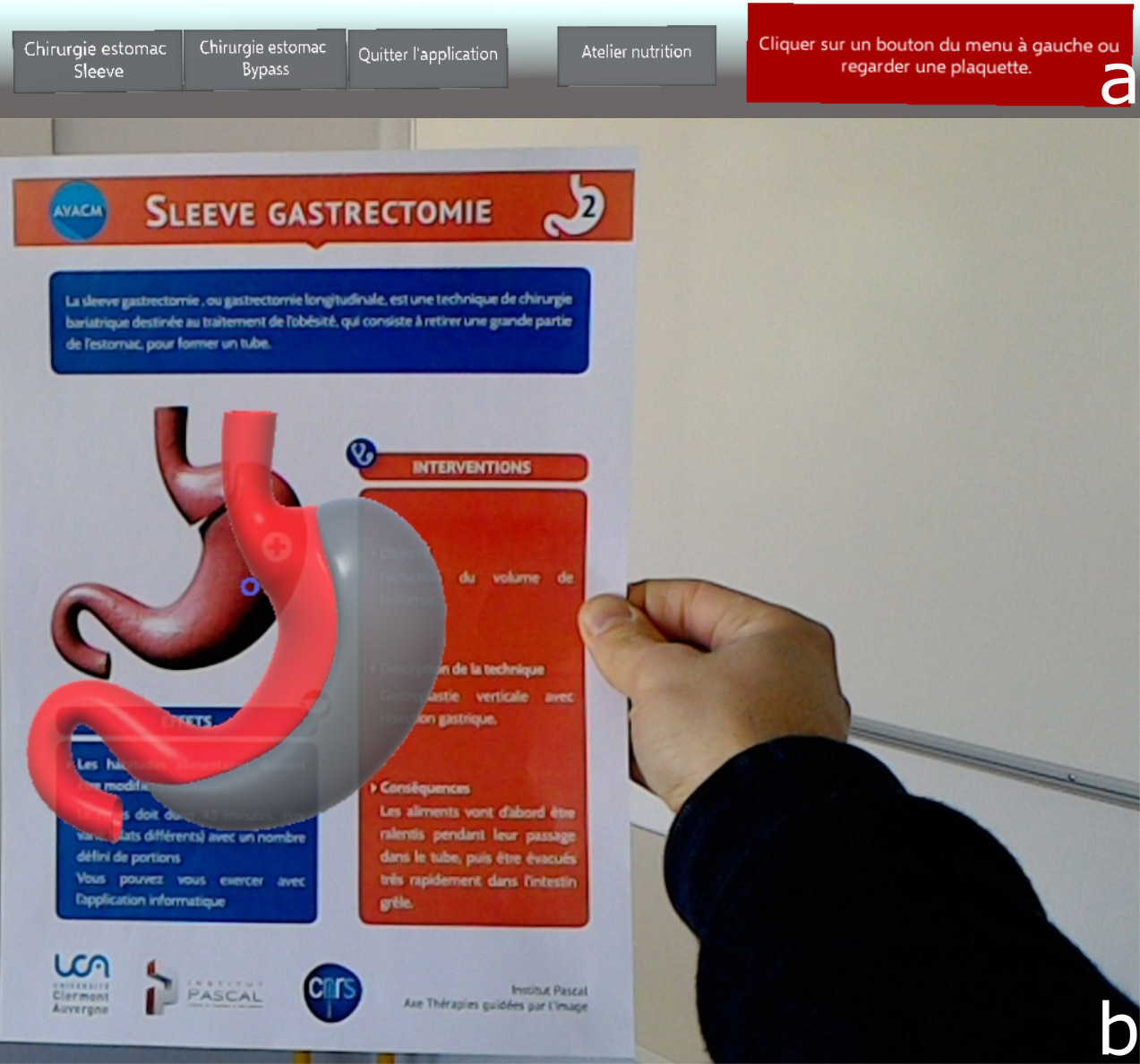}
	\caption{(a) Sleeve / Bypass Surgery toolbars. (b) Scanned leaflet (Sleeve intervention). A virtual 3D stomach occurs in front of the leaflet when the latter is scanned by the 
	headset.}
	\label{figure3}
\end{figure}

\begin{figure}[!t]	
	\centering
	\includegraphics[width=0.5\textwidth]{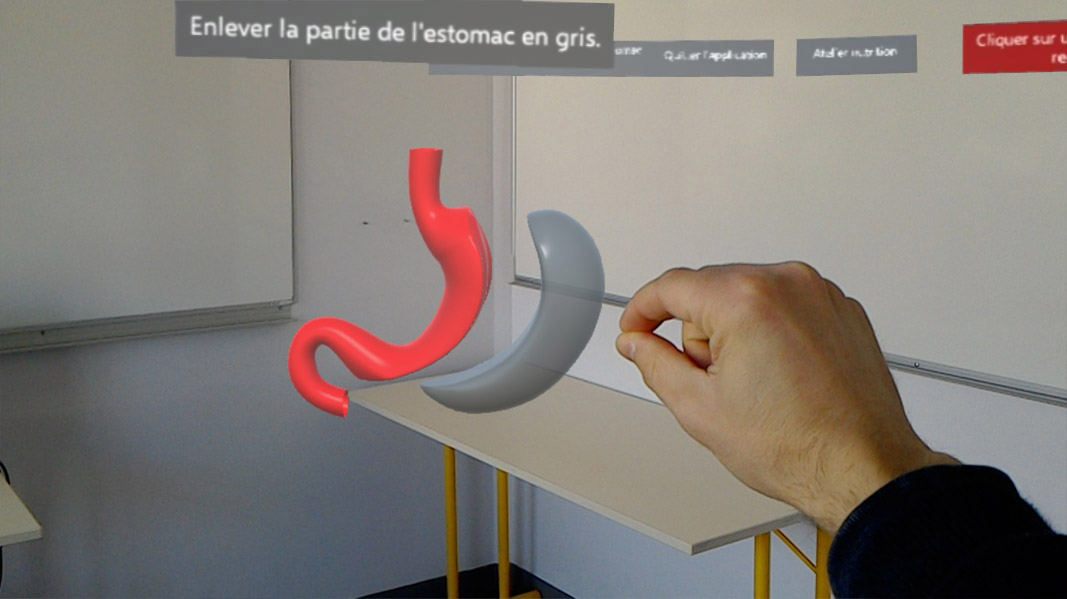}
	\caption{User interaction with the virtual stomach. Frame retrieved from \textbf{SEQ\_GESTURES} sequence (Section~\ref{subsecStep4 Presentation hardware}).}
	\label{figure4}
\end{figure}

 The first stage shows the user the type of intervention she or he will undergo (either using buttons in the application (Figure~\ref{figure3}(a)), or by scanning a leaflet 
with the HoloLens (Figure~\ref{figure3}(b)), knowing that there is a leaflet for each type of intervention. For the implementation of the communication that will be described later 
(sections 3 and 4), we use Photoshop and Illustrator from Adobe Creative Suite \cite{Adobe2018}. Then a virtual stomach is displayed on which the user will interact (for example, 
for the Sleeve operation, the user will select a portion of the stomach and remove it (Figure~\ref{figure4}). This analogous corresponding part was not present in the conventional 
workshop \textbf{CONV\_WORKSHOP}. Once the action is done, the second stage starts.

\begin{figure}[!t]	
	\centering
	\includegraphics[width=\textwidth]{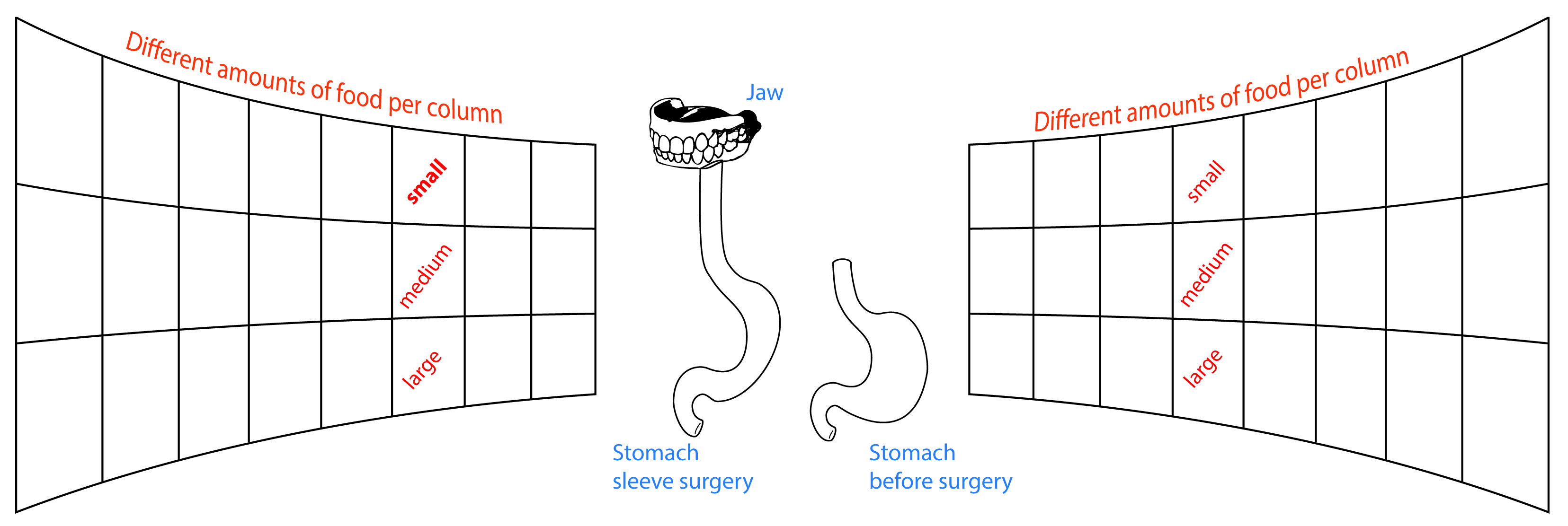}
	\caption{Self-service restaurant shown as a half cylinder model.}
	\label{figure5}
\end{figure}

The second stage shows a virtual self-service restaurant containing thirty foods. By the amount of information and so that the patient can interact easily, this self-service
restaurant is proposed as a half-cylinder model with three different quantities (small, medium, large) of a given food in the same column (Figure~\ref{figure5}). This seems to 
be the optimal solution to present a maximum of elements with which the user wearing the headset can interact. In other words, the environment has been specifically designed in 
relation to the use of such headsets.

These foods are illustrated as 2D images from the SUVIMAX study \cite{Hercberg2004,Hercberg2004b}. The user then chooses her or his food by looking at it (Gaze of the HoloLens) 
then grasps it by pinching her or his fingers and then releasing them (Gesture of the HoloLens). The food appears in front of a 3D jaw in the shape of a cube (Figure~\ref{figure5}). 
The user grasps the food again and brings it to her or his own mouth, an animation (cycle of opening and closing of the jaw) is then launched so that the user becomes aware of the chewing 
time. The user repeats this process as long as she or he thinks it is necessary to add food to compose her or his menu. When the menu composition is achieved, a first assessment is proposed 
concerning the number of meals and the corresponding quantity. Then the quantitative and qualitative messages, the same as those of \textbf{CONV\_WORKSHOP}, are displayed. The user 
can then repeat this training a second time. During this second iteration, additional indicators (number of portions ingested, \textit{etc.}) can help the patient to better compose 
her or his menu.

\leftskip=0.6cm
\subsubsection{Setting up the digital workshop \textbf{DIG\_WORKSHOP}}
\label{subsubsecsetDig}
\leftskip=0cm

This workshop was performed six times with three new patients each time who had to use \textbf{HOLO\_NUTRI} at the same time. Each patient wore a HoloLens headset. 
Since the headset is semi-transparent, patients could then keep in touch with the coordinating nurse. The solution is quite complex to set up:

\begin{itemize}
	\item this headset must be presented to patients who, for a few, have poor skills in computer science,
	\item the gestures to the HoloLens can be quite complicated to perform,
	\item the connection between the computer and the HoloLens can cause problems.
\end{itemize}

Unlike \textbf{CONV\_WORKSHOP}, a team of computer scientists is required to assist patients to use this headset and therefore this application. We then chose to display a 
visual feedback of what the user sees by connecting the headset to a laptop: each patient is then assisted by a computer specialist who can help her or him through the visual 
feedback on the computer screen.

\par Nevertheless the guidance of a patient by a computer scientist is not enough for the effective exploitation of the application. The latter can be described as "pure mixed 
reality" in the sense that it is not proposed to use only the traditional computer / screen / keyboard / mouse set; indeed, the objectives described in section~\ref{subsubsecWorkshopDig} would not be 
reached and in particular the gesture made by the patient to grab the food. Users may have difficulty using mixed reality, unlike a more traditional computer program. In order to 
guarantee a good flow, this application requires a very specific communication so that patients are as autonomous and efficient as possible.

\par We then describe the communication we have put in place for this workshop; it relies on the use of multimedia supports briefly described in section~\ref{secPrelimDIGWORKSHOP} and more precisely in 
section~\ref{secDescglobalcomm}. We will, first of all, describe the different stages of the workshop \textbf{DIG\_WORKSHOP} in section~\ref{secDIGWORKSHOP} so as to understand the proposed communication.  

\section{Description of the DIG\_WORKSHOP}
\label{secDIGWORKSHOP}

The workshop \textbf{DIG\_WORKSHOP} consists of 9 steps (see Table 2): Introduction of the session (step 1), identical to step 1 of \textbf{CONV\_WORKSHOP}; debate around the 
question "feeding ourself, eating, when we have been under a bariatric surgery, what does it bring to mind? (Step 2), identical to step 2 of \textbf{CONV\_WORKSHOP}; presentation of 
the group of computer scientists (step 3); presentation of the material and its use (step 4) through a video. We have developed a software \textbf{LEARN\_EX} to learn the basics of 
the HoloLens that are essential to the use of \textbf{HOLO\_NUTRI}. We then give a user manual on this learning exercise \textbf{LEARN\_EX} (step 5), exercice that patients must 
complete. Then, we begin the workshop to discuss with patients about their daily diet after intervention (step 6), identical to step 5 of \textbf{CONV\_WORKSHOP}; patients are invited to 
compose their menu within \textbf{HOLO\_NUTRI} software (step 7); the dietician then gives each patient advices based on the foods and quantities she or he has chosen (step 8); 
patients complete forms to evaluate the session (step 9), the same step in step 9 of \textbf{CONV\_WORKSHOP}.

\begin{table}
	\centering
	\caption{Description of both steps and actions during digital workshop \textbf{DIG\_WORKSHOP}. Actors of all actions are identified by specific colors detailed at the end of this table.}
	\label{table2}
	\setlongtables
	\begin{tabular}{|l|l|l|l|}
	\hline
	\multicolumn{4}{|l|}{1. Session introduction}  \\ 
	\hline
	 & \multicolumn{3}{|l|}{\cellcolor[HTML]{4f81bd}{\color[HTML]{ffffff}{1.1. Experimental protocol presentation and free consent form signature}}} \\ 
	\hline
	 & \multicolumn{3}{|l|}{\cellcolor[HTML]{9bbb59}1.2. Signature of the free consent form by each patient} \\
	\hline
	 & \multicolumn{3}{|l|}{\cellcolor[HTML]{fabf8f}1.3. Group presentation (clinicians/patients)} \\
	\hline
	 & \multicolumn{3}{|l|}{\cellcolor[HTML]{4f81bd}{\color[HTML]{ffffff}{1.4. Session objectives presentation}}} \\
	\hline
	\multicolumn{4}{|l|}{}  \\ 
	\hline
	\multicolumn{4}{|l|}{\cellcolor[HTML]{fabf8f}\begin{tabular}[c]{@{}l@{}}2. Debate around the question «Feeding ourselves, eating, when we have been \\ under a bariatric surgery, what does it bring to mind?»\end{tabular}}  \\ 
	\hline
	\multicolumn{4}{|l|}{}  \\ 
	\hline
	\multicolumn{4}{|l|}{\cellcolor[HTML]{00b0f0}{\begin{tabular}[c]{@{}l@{}}3. Group presentation (computer scientists/patients)\end{tabular}}} \\ 
	\hline
	\multicolumn{4}{|l|}{}  \\ 
	\hline
	\multicolumn{4}{|l|}{4. Hardware presentation}  \\ 
	\hline
	 & \multicolumn{3}{|l|}{\cellcolor[HTML]{c00000}{\color[HTML]{ffffff}{4.1. HoloLens presentation (oral + hardware)}}} \\ 
	\hline
	 & \multicolumn{3}{|l|}{\cellcolor[HTML]{c00000}{\color[HTML]{ffffff}{4.2. Headset presentation and its carrying (video sequence \textbf{SEQ\_HEADSET} + oral)}}} \\
	\hline
	 & \multicolumn{3}{|l|}{\cellcolor[HTML]{c00000}\color[HTML]{ffffff}{{4.3. Gesture presentation (video sequence \textbf{SEQ\_GESTURES} + oral)}}} \\
	\hline
	 & \multicolumn{3}{|l|}{\cellcolor[HTML]{c00000}{\color[HTML]{ffffff}{4.4. Reminder leaflet delivery (leaflet \textbf{LEAFLET\_HEADSET\_GESTURES})}}} \\
	\hline
	\multicolumn{4}{|l|}{}  \\ 
	\hline
	\multicolumn{4}{|l|}{5. Learning exercise}  \\ 
	\hline
	 & \multicolumn{3}{|l|}{\cellcolor[HTML]{c00000}{\color[HTML]{ffffff}{5.1. Learning exercise presentation (video sequence \textbf{SEQ\_LEARN\_EX} + oral)}}} \\ 
	\hline
	& \multicolumn{3}{|l|}{\cellcolor[HTML]{c00000}{\color[HTML]{ffffff}{5.2. Reminder leaflet delivery (1 \textbf{LEAFLET\_LEARN\_EX} per person)}}} \\ 
	\hline
	 & \multicolumn{3}{|l|}{\cellcolor[HTML]{00b0f0}{5.3. Headset setup (computer scientists/patients)}} \\
	\hline
	 & \multicolumn{3}{|l|}{\cellcolor[HTML]{9bbb59}{5.4. Launching of learning exercise \textbf{LEARN\_EX}}} \\
	\hline
	 & \multicolumn{3}{|l|}{\cellcolor[HTML]{9bbb59}{5.5. Headset removal}} \\
	\hline
	\multicolumn{4}{|l|}{}  \\ 
	\hline
	\multicolumn{4}{|l|}{6. Workshop trigger}  \\ 
	\hline
	 & \multicolumn{3}{|l|}{\cellcolor[HTML]{fabf8f}{6.1. Starting question «After the surgery, how to eat daily?»}} \\ 
	\hline
	\multicolumn{4}{|l|}{}  \\ 
	\hline
	\multicolumn{4}{|l|}{7. HOLO\_NUTRI application}  \\ 
	\hline
		 & \multicolumn{3}{l|}{\cellcolor[HTML]{c00000}{\color[HTML]{ffffff}{7.1. Presentation of \textbf{HOLO\_NUTRI} application (video sequence \textbf{SEQ\_HOLO}}}}\\
		 & \multicolumn{3}{l|}{\cellcolor[HTML]{c00000}{\color[HTML]{ffffff}{\textbf{\_NUTRI} + oral)}}}\\		 
		 & \multicolumn{3}{l|}{\cellcolor[HTML]{c00000}{\color[HTML]{ffffff}{7.2. Distribution of paper reminder on the application (1 leaflet \textbf{LEAFLET\_HOLO}}}}\\
		 & \multicolumn{3}{l|}{\cellcolor[HTML]{c00000}{\color[HTML]{ffffff}{\textbf{\_NUTRI} per person according to the type of surgical operation)}}}\\
		 & \multicolumn{3}{l|}{\cellcolor[HTML]{00b0f0}7.3. Headset setup (computer scientists/patients)}\\
		 & \multicolumn{3}{l|}{\cellcolor[HTML]{ffffff}7.4. Launching of \textbf{HOLO\_NUTRI}}\\
		 & \multicolumn{3}{l|}{\cellcolor[HTML]{9bbb59} \hspace*{0.5cm}7.4.1. Surgical operation simulation (using leaflet or the menu buttons)}\\
		 & \multicolumn{3}{l|}{\cellcolor[HTML]{ffffff} \hspace*{0.5cm}7.4.2. First iteration (simple interface) and second iteration (same interface}\\  
		 & \multicolumn{3}{l|}{\cellcolor[HTML]{ffffff} \hspace*{0.5cm}with more visual information and with continuous analysis of the food choice)}\\
		 & \multicolumn{3}{l|}{\cellcolor[HTML]{9bbb59} \hspace*{1cm}7.4.2.1. Food choice via 2D images in the virtual self-service restaurant}\\
		 & \multicolumn{3}{l|}{\cellcolor[HTML]{bfbfbf} \hspace*{1cm}7.4.2.2. Recap panel}\\
		 & \multicolumn{3}{l|}{\cellcolor[HTML]{bfbfbf} \hspace*{1cm}7.4.2.3. Analyse of the patient choice}\\
		 & \multicolumn{3}{l|}{\cellcolor[HTML]{bfbfbf} \hspace*{1cm}7.4.2.4. Informative messages}\\
		 & \multicolumn{3}{l|}{\cellcolor[HTML]{9bbb59}7.5. Headset removal}\\
	\hline
	\multicolumn{4}{|l|}{}  \\ 
	\hline
	\multicolumn{4}{|l|}{\cellcolor[HTML]{fabf8f}{\begin{tabular}[c]{@{}l@{}}8. Discussion on each patient choice in an individual manner\end{tabular}}} \\  
	\hline
	\multicolumn{4}{|l|}{}  \\ 
	\hline
	\multicolumn{4}{|l|}{\cellcolor[HTML]{9bbb59}{\begin{tabular}[c]{@{}l@{}}9. Forms (knowledge evaluation + satisfaction)\end{tabular}}} \\ 
	\hline
	\end{tabular}
	\end{table}

\begin{table}
\begin{tabular}{llllll}
\cellcolor[HTML]{4f81bd} & \textit{\begin{tabular}[c]{@{}l@{}}Clinical team\\ action\end{tabular}} & \cellcolor[HTML]{fabf8f} & \textit{\begin{tabular}[c]{@{}l@{}}Clinical team + patients\\ action\end{tabular}} & \cellcolor[HTML]{9bbb59} & \textit{\begin{tabular}[c]{@{}l@{}}Patients\\ action\end{tabular}} \\
                         &                                                        &                          &   &                          &   \\
\cellcolor[HTML]{c00000} & \textit{\begin{tabular}[c]{@{}l@{}}Computer scientists\\ action\end{tabular}} & \cellcolor[HTML]{00b0f0} & \textit{\begin{tabular}[c]{@{}l@{}}Computer scientists team + patients\\ action\end{tabular}} & \cellcolor[HTML]{bfbfbf} & \textit{\begin{tabular}[c]{@{}l@{}}Application\\ action\end{tabular}} \\
\end{tabular}
\end{table}

\section{Preliminary thinking on the setting of a communication dedicated to the digital workshop DIG\_WORKSHOP}
\label{secPrelimDIGWORKSHOP}

From a logistical point of view, the constraints we face for this kind of workshop are multiple (small room space, several patients at the same time, time limit of 1:30 hour for a
3-person group who must stay seated for the entire session). The official tutorial offered by Microsoft aims at computer experienced users. Here, we have to present the headset as 
simply and efficiently as possible in order to this one to be used in the most optimal way no matter the skills the user could have in computer science. It was also unnecessary to 
show all features of this headset as they are not all used in the \textbf{HOLO\_NUTRI} application (\textit{e.g.} voice control). In other words, we could not use original demonstrations
as proposed by the headset construtor, we had to develop a dedicated communication designed both for patients and for the purpose of our application.     

\subsection{Global communication process}
\label{subsecGlobal}
We provide all communication actions designed specifically for \textbf{DIG\_WORKSHOP} by referring to Table 2:

\begin{itemize}
	\item Step 1. Presentation by the usual team (clinicians) of the workshop objectives (awareness on feeding modification after the surgery), all workshop steps presentation and 
	free consent form signature,
	\item Step 3. Computer scientists team presentation,
	\item Step 4. Hardware presentation:
	\begin{itemize}
		\item Step 4.1. HoloLens headset presentation,
		\item Step 4.2. Presentation of the video sequence \textbf{SEQ\_HEADSET} explaining the headset carrying,
		\item Step 4.3. Presentation of the video sequence \textbf{SEQ\_GESTURES}  showing the gestures to perform,
		\item Step 4.4. Hanging out of a reminder (leaflet \textbf{LEAFLET\_HEADSET\_GESTURES}) summarizing steps 4.2 and 4.3,
	\end{itemize}
	\item Step 5. Learning exercise \textbf{LEARN\_EX} for practice with the headset (gaze direction and gestures):
	\begin{itemize}
		\item Step 5.1. Presentation of the video sequence \textbf{SEQ\_LEARN\_EX} showing the learning exercise \textbf{LEARN\_EX},
		\item Step 5.2. Delivery of a reminder (leaflet \textbf{LEAFLET\_LEARN\_EX}) summarizing step 5.1,
	\end{itemize}
	\item Step 7. \textbf{HOLO\_NUTRI} application:
	\begin{itemize}
		\item Step 7.1. Presentation of the video sequence \textbf{SEQ\_HOLO\_NUTRI},
		\item Step 7.2. Hanging out the leaflet \textbf{LEAFLET\_HOLO\_NUTRI} in accordance to the considered operation type, and allowing its scan while \textbf{HOLO\_NUTRI} 
		application running,
	\end{itemize}
	\item Step 9. Feedback on this experience by answering a form.
\end{itemize}

The communication is described in more details for each of those steps in the next section.

\section{Description of the content of the global communication}
\label{secDescglobalcomm}

In this section, we detail the different steps proposed in the previous section, while putting forward the differences of communication between the conventional session 
\textbf{CONV\_WORKSHOP} and the digital session \\\textbf{DIG\_WORKSHOP} resulting from the exploitation of the mixed reality by patients. 
In addition to reviewing these different steps, we justify the choices of communication or used media as well as their design to be usable both in the digital workshop \textbf{DIG\_WORKSHOP}
and in practical by a HoloLens headset. 
More particularly, we will thus detail the content of the learning exercise (content) designed to understand the gestures to handle HoloLens headset (section~\ref{subsecDescriptLEARNEX}), the video sequences 
(sections~\ref{subsubsecSEQ} and \ref{subsubsubsecSEQHOLO}) and also the design of the leaflets (content and conception) (section~\ref{subsubsecLEAFLET} and \ref{subsubsubsecDescripLeaflet}). 
The reflections are certainly constrained within the framework of our workshop, but could be useful to other people in a completely different framework of exploitation of mixed reality.

\subsection{Description of steps 1 to 4}
\label{subsecSteps1a4}
\leftskip=0.6cm
\subsubsection{Step 1. Session introduction}
\label{subsecStep1Sessionintro}
\leftskip=0cm

\begin{itemize}
\item Steps 1.1 and 1.2. Presentation of the protocol by the clinical team and signature of the consent forms by the patients (participation in the study, image rights),
\item Step 1.3 Presentation of the group of clinician and patients,
\item Step 1.4 Presentation of the objectives of the session.
\end{itemize}

\leftskip=0.6cm
\subsubsection{Step 3. Presentation of the team of computer scientists.}
\label{subsecStep3 Presentation computer scientist}
\leftskip=0cm

\begin{figure}[!t]	
	\centering
	\includegraphics[width=1\textwidth]{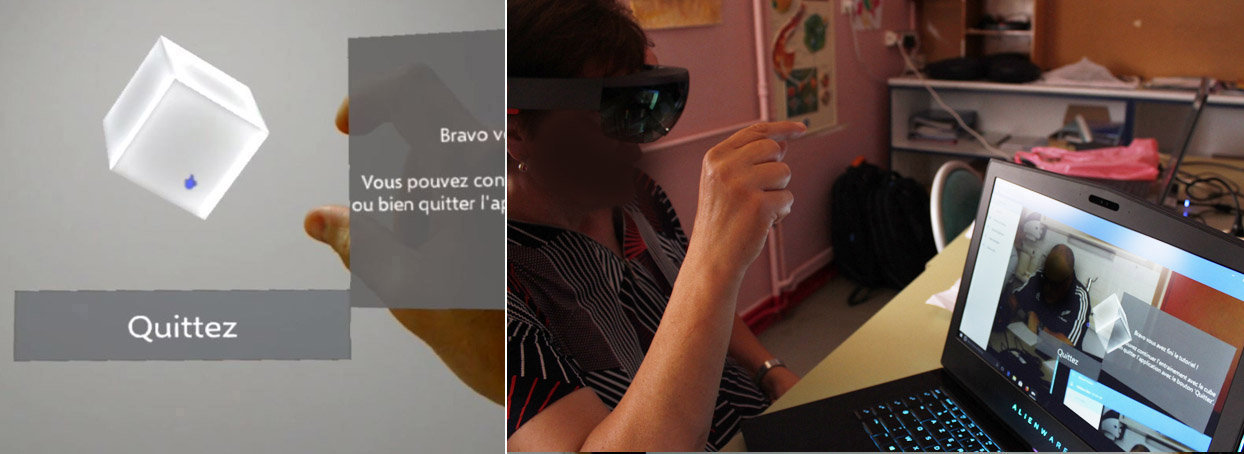}
	\caption{Visual feedback (seen by a patient through the headset (left), shown on the laptop screen (right)).}
	\label{figure6}
\end{figure}

\begin{itemize}
\item The three patients are installed around an oval table, a laptop is installed in front of each patient. A computer scientist is located on the side of each patient 
and can then see the visual feedback of what the patient sees in the headset, on the laptop screen (Figure~\ref{figure6}).
\end{itemize}

\leftskip=0.6cm
\subsubsection{Step 4. Presentation of the hardware}
\label{subsecStep4 Presentation hardware}
\leftskip=0cm

\begin{figure}[!t]	
	\centering
	\includegraphics[width=0.5\textwidth]{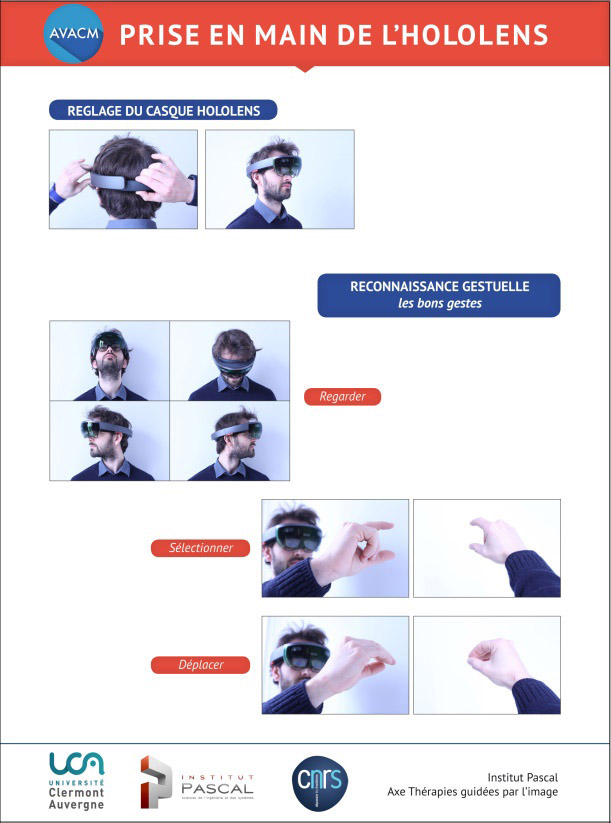}
	\caption{\textbf{LEAFLET\_HEADSET\_GESTURES} leaflet.}
	\label{figure7}
\end{figure}

\begin{itemize}
\item Step 4.1. Oral presentation of the HoloLens headset. A computer scientist shows the headset to the patients so that they realize the weight and the bulk of the hardware. 
Some notions of mixed reality are introduced: possibility to see the real environment and 3D objects on the transparent visor of the headset, taking into account the space around 
the patient for the display and manipulation of virtual objects. This step also aims to reassure patients who are not necessarily used to computers and even less devices like these 
headsets.
\item Step 4.2. Presentation of the video sequence \textbf{SEQ\_HEADSET} about the headset and its carrying. It is simultaneously commented by computer scientists. 
The duration of this sequence is 52 seconds.

\par Description of the video sequence \textbf{SEQ\_HEADSET}: an operator shows the headset from different views (front, rear, side); then the operator positions 
the headset on his head and shows that he can sweep his environment by turning his head. During the sequence, we indicate that it is the orientation of the head that is taken into 
account by the HoloLens and that the headset does not allow the follow-up of the eyes (known as \textit{eye-tracking}; note that HoloLens in its second version, not yet available during our 
experiment, takes into account eye-tracking).

\item Step 4.3. Presentation of the video sequence \textbf{SEQ\_GESTURES} on gestures recognized by the HoloLens, specifically only the gestures used in the \textbf{LEARN\_EX} 
learning exercise and in the \textbf{HOLO\_NUTRI} application are presented, these gestures are identical for these two applications. The duration of this sequence is 57 seconds.
\par The interaction system based on the gesture recognition is introduced. Different gestures are natively recognized by the hardware to interact, that is to say a precise gesture 
(or a combination of gestures) is associated with a specific action. For example, pinching the thumb and forefinger with the rest of the hand closed triggers the “selection action”, 
releasing the thumb and index finger after the selection results in the “validation action”, holding the inch and index finger triggers the “move action”.

\par In this sequence \textbf{SEQ\_GESTURES}, we describe these three gestures and highlight on the good positioning of the hand to perform these gestures (Figure~\ref{figure4}). 
We also show that it is possible to do wrong gestures (for example, selection and pinching errors) always in order to reassure patients and not to rush during the experiment. 
We deem that the part presenting the bad handling to make the gestures (too fast speed of pinching, bad orientation of the hand) is mandatory. During pauses occurring while the video 
is playing, we also highlighted the presence of cameras in the headset which detect hands: this technical justification implies that gestures must be made correctly (right angle 
between the index finger and thumb to select, pinch speed, \textit{etc.}) and should be done according to the orientation of the head. The announcement of this technical constraint 
made it easier for patients to correct themselves, and thus to perform the session more efficiently.

\item Step 4.4. Distribution of the leaflet \textbf{LEAFLET\_HEADSET\_GESTURES}

\par The team of computer scientists distributes \textbf{LEAFLET\_HEADSET\_GESTURES} leaflet (Figure~\ref{figure7}) containing the information of the two video sequences 
(\textbf{SEQ\_HEADSET} and 
\textbf{SEQ\_GESTURES}), namely the setting of the port of the headset (step 4.2) as well as a summary of the basic gestures to be carried out (step 4.3): orient your head, point 
your hand towards the headset camera, spread your thumb and forefinger. Moreover, thanks to the transparent visor of the HoloLens, the patient can at any time visualize the most 
important steps of the application included on this leaflet, provided that the occupation of the information projected on the visor leaves enough of space in the user's field of 
view.
\end{itemize}

\subsection{Step 5. Learning exercise \textbf{LEARN\_EX}}
\label{subsecLEARNEX}

To successfully use the application \textbf{HOLO\_NUTRI}, it is mandatory that patients gestures are performed correctly with the HoloLens. It quickly became necessary to offer a 
learning exercise to patients, and as noted above, the most suitable for the application \textbf{HOLO\_NUTRI} (and not to use the official tutorial). We first describe the 
\textbf{LEARN\_EX} learning exercise and then the associated communication.

\leftskip=0.6cm
\subsubsection{Description of the learning exercise \textbf{LEARN\_EX}}
\label{subsecDescriptLEARNEX}
\leftskip=0cm
 
The learning exercise \textbf{LEARN\_EX} (Figure~\ref{figure8}) is composed of the following three stages: 

\begin{figure}[!t]	
	\centering
	\includegraphics[width=\textwidth]{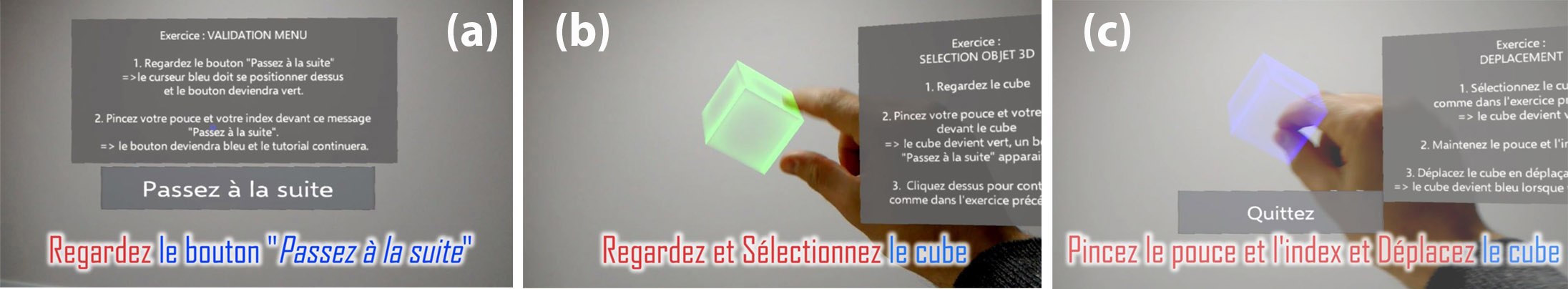}
	\caption{The three stages of the learning exercise \textbf{LEARN\_EX}: actions (a) Menu validation, (b) 3D object selection, (c) 3D object displacement. Frames extracted from our video sequence \textbf{SEQ\_LEARN\_EX}.}
	\label{figure8}
\end{figure}

\begin{itemize}
\item \textbf{Stage 1: Menu Validation Action} (Figure~\ref{figure8}(a)) > This step teaches patients to look at a button (whose default color is gray). When the gaze is put on the 
button after a few seconds, the button turns green. The patient can then select it by performing the "pinching" gesture, then the "loosening" gesture. Once the gesture is done correctly, the button displays the text to the next step and turns
blue.
\item \textbf{Stage 2: 3D Object Selection Action} (Figure~\ref{figure8}(b)) > In this step, patients learn to look at a 3D object, a gray cube. Like stage 1, the cube turns green 
after a few seconds when the gaze is held on the button. The "pinching" gesture allows the user to select the cube, the "loosening" gesture to deselect the cube and to display a next 
button to move forward, on that button you have to repeat the same gestures as during stage 1 to go through in the next stage.
\item \textbf{Stage 3: Move Action} (Figure~\ref{figure8}(c)) > This last stage teaches patients to look at the cube and with the only "pinching" gesture, \textit{i.e.} keeping the thumb and 
index finger tight, thus the cube becomes blue. It is then possible to shift it by the gesture "movement of the arm". The displacement stops during the "loosening" gesture. Then the 
learning exercise stops.
\end{itemize}

The objective of \textbf{LEARN\_EX} is that the user becomes familiar with this new type of manipulation. Unlike the official tutorial provided with the headset 
(heavy digital scenes), we propose an application manipulating only a 2D object (menu button), a 3D object (cube), the 3 actions (validate, select, displace) and the associated 
gestures or gesture combinations (pinching, loosening and moving the arm). This allows patients to be as effective as possible on this new technology for our purposes.

\par Therefore the user becomes more familiar, on one hand, with 3D objects (holograms), her or his environment and this new augmented visualization both, and on the other hand, with 
their selection and their displacement in the 3D space. The positioning of the gaze and the 3 actions "validation", "selection" and "displacement", by the gestures “pinching”, 
“loosening” and “movement of the arm” are also the only actions and gestures necessary to perform the application \textbf{HOLO\_NUTRI}. We also took on the same visual appearance on 
the graphical elements of the interface for the two “validation” and “selection” actions between the \textbf{LEARN\_EX} learning exercise and the \textbf{HOLO\_NUTRI} application.

The communication for \textbf{LEARN\_EX} is composed of two supports. The first one, described in section~\ref{subsubsecSEQ}, is the \textbf{SEQ\_LEARN\_EX} video sequence explaining the 
exercise (step 5.1). The second one, described in section~\ref{subsubsecLEAFLET}, is the leaflet \textbf{LEAFLET\_LEARN\_EX}. The latter is given out to patients during the step 5.2.
We now detail these two supports.

\leftskip=0.6cm
\subsubsection{Step 5.1. Description of the video sequence \textbf{SEQ\_LEARN\_EX} of the learning exercise \textbf{LEARN\_EX}.}
\label{subsubsecSEQ}
\leftskip=0cm

The duration of the video sequence (Figure~\ref{figure8}) is of 1 minute and 13 seconds. We chose to show the whole sequence to patients so that they do not have to put on and take 
off the headset between each stage of the exercise \textbf{LEARN\_EX}. In addition, it would have been difficult for a user to view this video clip played thoroughly on a computer 
screen, through the visor of the headset.

\par
\begin{figure}[!t]	
	\centering
	\includegraphics[width=\textwidth]{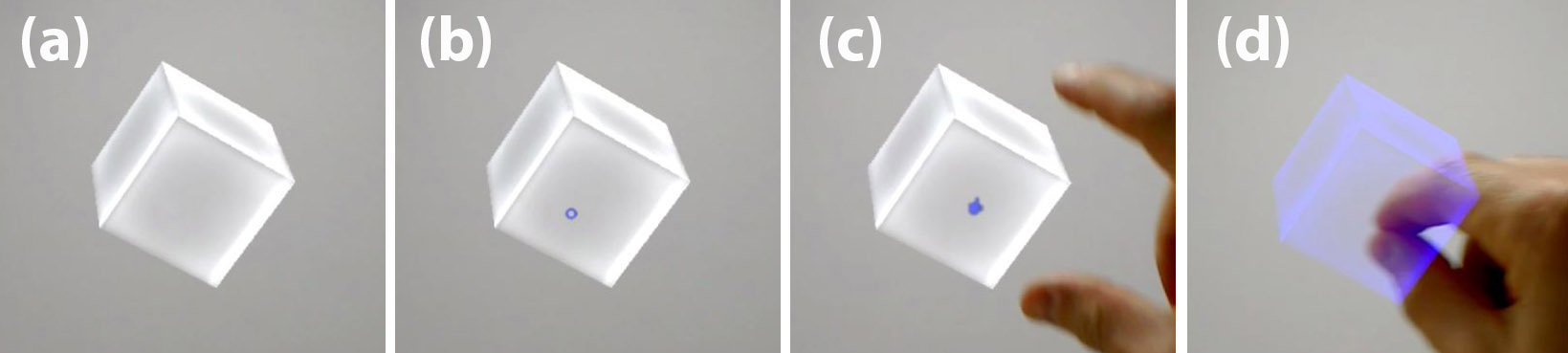}
	\caption{Display and cursor shape: (a) cube, (b) reticle, (c) hand shape, (d) move action.}
	\label{figure9}
\end{figure}

 During the viewing of the sequence, we emphasize the importance of the gaze when selecting an object in the environment. Take the example of a cube (Figure~\ref{figure9}(a)). 
When we look at the cube, a cursor occurs with the form of a disc (reticle) (Figure~\ref{figure9}(b)). We recall that the movement of the eyes is not taken into account by this 
headset; for the sake of simplicity, we call “put the gaze on” the "orientation of the head" action. Thus, any interaction with the environment is carried out in two stages: the user 
must first pose and maintain his gaze on a hologram (3D object or button to be selected), then perform the expected gesture while maintaining the gaze on the object. When the gaze is 
putted on the object of interest, then the reticle is transformed into a hand shape after a few seconds (Figure~\ref{figure9}(c)) if the camera has detected that the hand was ready 
to perform one of the expected actions on this object (corresponding to an action programmed in the application); it is then possible to interact (Figure~\ref{figure9}(d)) with this 
object (selection, validation, displacement).

Note that the video seems to be the proper support for patients to understand the notions of reticle, its transformation in a hand shape (indicating that a gesture is ready to be 
detected on the object). We realized that the biggest difficulties with the learning exercise were:

\begin{itemize}
\item to focus on the object of interest, more specifically patients must understand that it is necessary to turn the head and not the eyes, 
\item to position one hand in front of the cameras of the headset, according to the orientation of the head,
\item to spread the index finger and the thumb to make the “pinching” gesture.
\end{itemize}

\leftskip=0.6cm
\subsubsection{Step 5.2 : Description of the leaflet \textbf{LEAFLET\_LEARN\_EX} of the learning exercise}
\label{subsubsecLEAFLET}
\leftskip=0cm

\begin{figure}[!t]	
	\centering
	\includegraphics[width=0.5\textwidth]{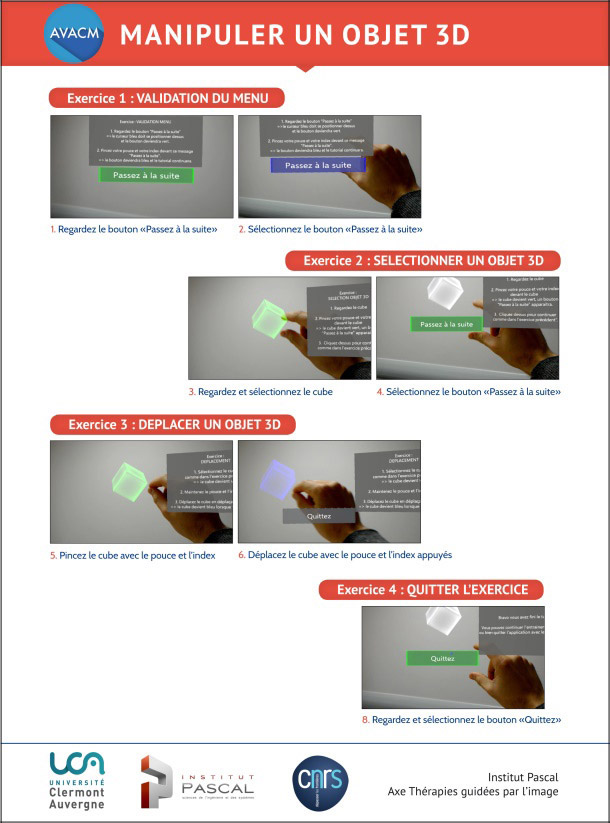}
	\caption{Leaflet \textbf{LEAFLET\_LEARN\_EX} summarizing the learning exercise \textbf{LEARN\_EX}.}
	\label{figure10}
\end{figure}

The leaflet \textbf{LEAFLET\_LEARN\_EX} (Figure~\ref{figure10}) recalls the gestures associated with the three actions: selection, validation and displacement (step 5.1).

\subsection{Step 7 : Nutrition session \textbf{HOLO\_NUTRI}}
\label{subsecNutri}
\leftskip=0.6cm
\subsubsection{Step 7.1 : Presentation of the video sequence \textbf{SEQ\_HOLO\_NUTRI} describing \textbf{HOLO\_NUTRI} application}
\label{subsubsubsecSEQHOLO}
\leftskip=0cm

We project the video sequence \textbf{SEQ\_HOLO\_NUTRI}, lasting 2 minutes and 58 seconds, presenting the contents arranged in the environment of the user. In the video, we emphasized the following points:

\begin{itemize}
\item the user must turn the head to scan the cylindrical virtual restaurant as a whole since the number of foods occupies a large visual space (Figure~\ref{figure2}, Figure~\ref{figure5} and Figure~\ref{figure11}).

\begin{figure}[!t]	
	\centering
	\includegraphics[width=\textwidth]{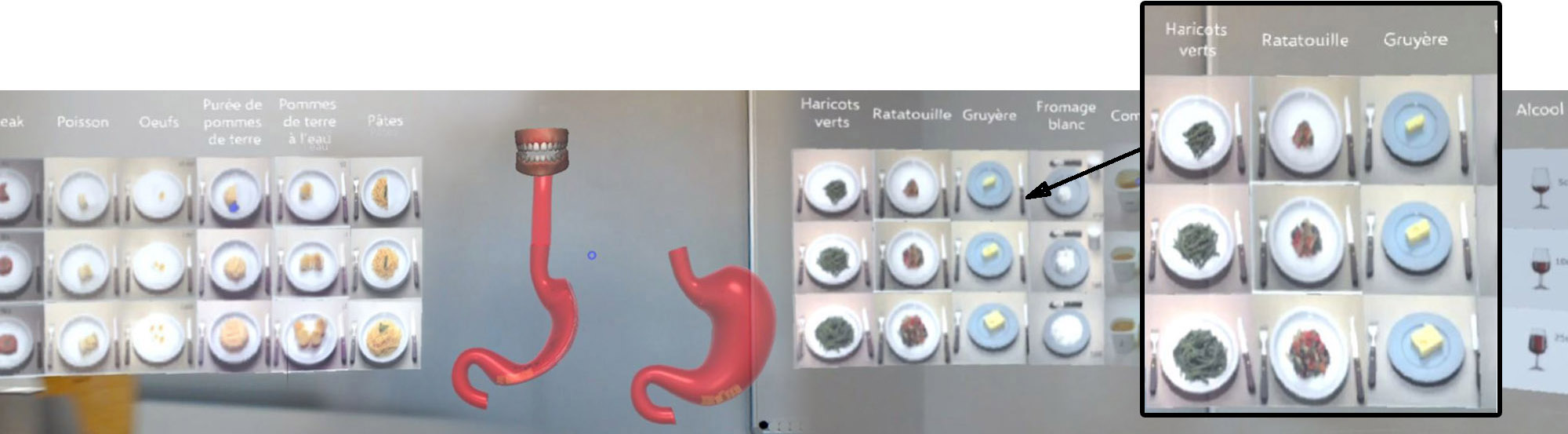}
	\caption{Screenshot of a part of the virtual self-service restaurant. Zoom-in boxed shows three different foods with three different quantities. Frame extracted from \textbf{SEQ\_HOLO\_NUTRI}.}
	\label{figure11}
\end{figure}

\item we inform users that different quantities of the same food are organized by rows (Figure~\ref{figure2}, Figure~\ref{figure5} and Figure~\ref{figure11}). 
\item we explain to the user that he must choose a food for a desired quantity. Once it is chosen, the food first appears in front of the virtual jaw, then the user must bring this food to her or his mouth. 
The gesture was not proposed in one go "bring the food to the mouth" but was broken down into two parts: to grab the food then to swallow it. Then a cycle of animation of the jaw starts preventing the user 
from taking another food until a predefined time (chewing time) has elapsed. This time is displayed in a timer,
\item we recall that there are two iterations and that in the second one, additional indicators are shown (stomach in transparency, color of the stomach according to the number of dishes and portions as well as 
indicator on the number of dishes and portions),
\item a summary is displayed with qualitative and quantitative messages corresponding to the menu composition.
\end{itemize}

\leftskip=0.6cm
\subsubsection{Step 7.2 : Distribution of the leaflet \textbf{LEAFLET\_HOLO\_NUTRI}}
\label{subsubsubsecLEAFLETHOLO}
\leftskip=0cm

We chose to have an introductory leaflet for each patient (step 7.2) from the beginning of the session in connection with the type of intervention she or he will undergo, as explained above. This leaflet contains 
both textual elements and an image representing one of the two types of interventions (Figure~\ref{figure14}).

\leftskip=0.6cm
\subsubsection{Interest of the leaflet for the \textbf{HOLO\_NUTRI application}}
\label{subsubsubsecHOLONUTRI}
\leftskip=0cm

The computer scientist assist the patient to launch the \textbf{HOLO\_NUTRI} application in the headset. The patient chooses the leaflet corresponding to her or his intervention, then puts the headset. The image of 
the stomach is scanned by the headset when the user looks it. A virtual 3D stomach appears above the leaflet, stomach having undergone the type of intervention determined by the leaflet (Bypass or Sleeve), as shown 
in Figure~\ref{figure3}. Then the user is asked to select a part of the stomach and to displace it (step 7.4.1), as shown in Figure~\ref{figure4}, the goal is that the patient realizes the shrinking of the stomach’s 
volume. Once the part of the stomach is removed, a new scene is launched in which the user can compose the menu; these are the two iterations of step 7.4.2.

\leftskip=0.6cm
\subsubsection{Description of the leaflet \textbf{LEAFLET\_HOLO\_NUTRI}}
\label{subsubsubsecDescripLeaflet}
\leftskip=0cm

Unlike the other two leaflets \textbf{LEAFLET\_HEADSET\_GESTURES} and \textbf{LEAFLET\_LEARN\_EX} that rather act as memory aid and reassure the patient, the leaflet 
\textbf{LEAFLET\_HOLO\_NUTRI} is scanned by the HoloLens headset: an adequate design had then to be proposed.

\par In this section, we describe the content of the leaflet and its layout. Then, we insist on the 
constraints related to HoloLens for the design of graphic elements of the leaflet (size and contrast of the image to be scanned). The two leaflets for chirurgical procedures, 
Bypass and Sleeve, are also shown.

\leftskip=0.6cm
\paragraph{Graphical content of the leaflet}
\label{subsubsubsecGraphic}
\leftskip=0cm

\begin{figure}[!t]	
	\centering
	\includegraphics[width=0.5\textwidth]{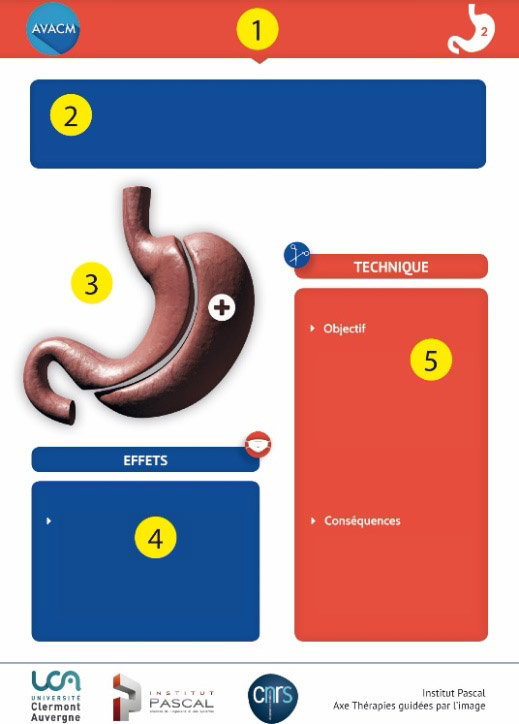}
	\caption{Template leaflet for a chirurgical procedure.}
	\label{figure12}
\end{figure}

The content of the leaflet (shown in Figure~\ref{figure12}) are : \textbf{\textit{Disk 1.}} Title: type of intervention for a given pathology, \textbf{\textit{Disk 2.}} Description of the 
pathology, \textbf{\textit{Disk 3.}} 3D organ: this image is a marker registered in the application database. When this marker is scanned by the HoloLens, a 3D stomach hologram 
appears allowing the simulation to start (step 7.4.1), \textbf{\textit{Disk 4.}} Brief explanation of eating habits, \textbf{\textit{Disk 5.}} Brief explanation of the surgical 
procedure.

\leftskip=0.6cm
\paragraph{Constraints related to HoloLens}
\label{subsubsubsecContraints}
\leftskip=0cm

\begin{figure}[!t]	
	\centering
	\includegraphics[width=\textwidth]{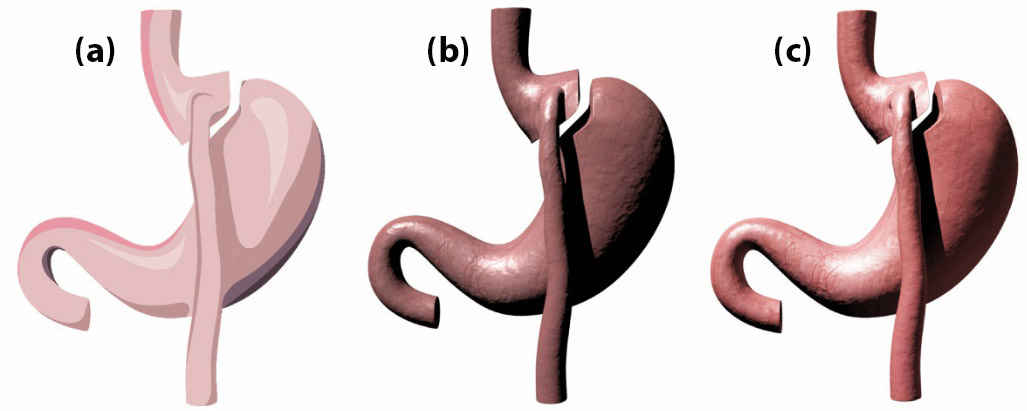}
	\caption{Marker tests for better detection by HoloLens: (a) flat design rendering, (b) initial 3D rendering, (c) 3D contrast rendering.}
	\label{figure13}
\end{figure}

We did several tests for marker detection by HoloLens in step 7.4.1, using the Vuforia library \cite{Vuforia2020}, directly integrated with the latest versions of Unity framework 
\cite{Unity2020}, that we use to develop application \textbf{HOLO\_NUTRI}. The first marker was in line with the visual identity in flat design (Figure~\ref{figure13}(a)) of the 
template leaflet described in the previous paragraph (Figure~\ref{figure12}), but the image was not recognized. We then proposed a marker with a volume feature 
(Figure~\ref{figure13}(b)), the marker was not recognized either. The detection finally works only by contrasting the previous marker (Figure~\ref{figure13}(c)).

\leftskip=0.6cm
\paragraph{Leaflets for both types of interventions}
\label{subsubsubsecLeaflets}
\leftskip=0cm

\begin{figure}[!t]	
	\centering
	\includegraphics[width=\textwidth]{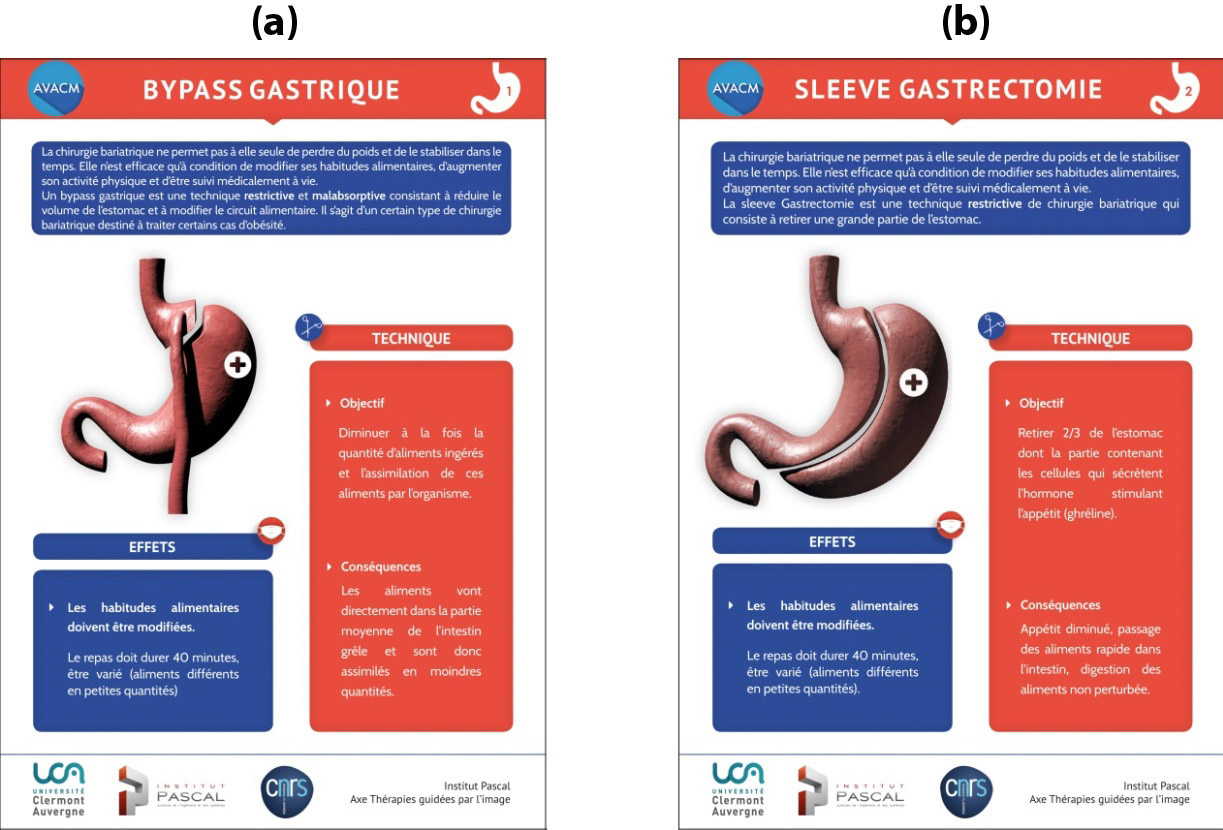}
	\caption{Leaflets \textbf{LEAFLET\_HOLO\_NUTRI} scanned by \textbf{HOLO\_NUTRI} (step 7.4.1): (a) Bypass, (b) Sleeve.}
	\label{figure14}
\end{figure}

The leaflet must be rigid enough both to be easily held in hand and durable to repeat the experience several times. Therefore we chose to laminate this leaflet. Laminating makes the leaflet brighter but does not 
alter the HoloLens' recognition of the leaflet. It should be noted that in case of unfavorable lighting conditions that do not allow the image to be scanned, the simulation (step 7.4.1) can also be launched by two 
menu buttons (Figure~\ref{figure3}(a)). The two resulting leaflets are shown in Figure~\ref{figure14}.

\section{Results}
\label{secResults}
We first report the duration of the video sequences and explanations projected and given to patients, then analyze the patient's assessment of different communication elements.

\subsection{Time information}
\label{subsecTimeinfo}

The sequences \textbf{SEQ\_HEADSET}, \textbf{SEQ\_GESTURES}, \textbf{SEQ\_LEARN\_EX} and \textbf{SEQ\_HOLO\_NUTRI} last respectively: 0:52 minute, 0:57 minute, 1:13 minute and 2:58 minutes. Ten sessions occurred at 
the hospital. The average watch time of the sequences \textbf{SEQ\_HEADSET} and \textbf{SEQ\_GESTURES} paused several times to give additional explanations (see section~\ref{subsecSteps1a4}) is 5 minutes (standard deviation 0:50 
minutes). The average duration of viewing of the \textbf{SEQ\_LEARN\_EX} sequence and the completion of the \textbf{LEARN\_EX} learning exercise is 4 minutes (standard deviation 2:38 minutes). The average duration 
of viewing of the sequence \textbf{SEQ\_HOLO\_NUTRI} is 4 minutes (standard deviation 00:42 minutes). To give a comparison order, the realization of the two iterations of \textbf{HOLO\_NUTRI} lasted on average 18 
minutes (standard deviation 5:54 minutes) \cite{Rositi2020}.

\subsection{Detail of the feasibility study following \textbf{LEARN\_EX}}
\label{subsecFeasibility}

Thirty patients participated at the sessions. The patients’ feedback following the completion of the \textbf{LEARN\_EX} learning exercise is given in Tables~\ref{table3} and \ref{table4}.

\begin{table}
	\caption{Patients’ feedback following the \textbf{LEARN\_EX} learning exercise (30 patients) - summary}
	\centering
	\begin{tabular}{|p{2.5cm}|p{1.5cm}|p{1cm}|p{1cm}|p{1cm}|p{1cm}|c|p{0.5cm}|}
	\hline	
		&     & Not at all & Not really & Yes, rather & Yes absolutely & NA\\ \hline
		Q1 After watching the video, do you think this app is easy to use ?& Number of persons & 0 & 8 & 16 & 6 & 0\\ \cline{2-7}
		& \% (on 30) & 0\% & 26.7\% & 53.3\% & 20.0\% & 0\% \\ \hline
		Q2 Did the introduction to the technical gestures of the application seems clear to you ?& Number of persons & 0 & 2 & 17 & 11 & 0\\ \cline{2-7}
		& \% (on 30) & 0\% & 6.7\% & 56.7\% & 36.6\% & 0\% \\ \hline
		Q3 Are you comfortable using this application ?& Number of persons & 2 & 9 & 12 & 7 & 0\\ \cline{2-7}
		& \% (on 30) & 6.7\% & 30.0\% & 40.0\% & 23.3\% & 0\% \\ \hline
	\end{tabular}
	\label{table3}
\end{table}

\begin{table}
	\caption{Patients’ feedback following the \textbf{LEARN\_EX} learning exercise (30 patients) – detail per patient}
	\centering
	\begin{tabular}{|p{2cm}|p{2.5cm}|p{2.5cm}|p{2.5cm}|}
	\hline	
	& Q1 & Q2 & Q3\\ \hline
	 Patient ID. & After watching the video, do you think this app is easy to use ? & Did the introduction to the technical gestures of the application seems clear to you ? & Are you comfortable using this application ?\\ \hline
	1 & yes absolutely & yes absolutely & yes absolutely\\ \hline
	2 & yes absolutely & \cellcolor[HTML]{dad500} yes, rather & \cellcolor[HTML]{ff0000} not at all\\ \hline
	3 & \cellcolor[HTML]{fea402} not really & \cellcolor[HTML]{dad500} yes, rather & \cellcolor[HTML]{fea402} not really\\ \hline
	4 & yes, rather & \cellcolor[HTML]{dad500} yes, rather & \cellcolor[HTML]{fea402} not really\\ \hline
	5 & yes absolutely & yes, rather & yes, rather\\ \hline
	6 & \cellcolor[HTML]{b2a1c7} not really & yes, rather & yes, rather\\ \hline
	7 & yes, rather & yes, rather & yes, rather\\ \hline
	8 & yes, rather & yes, rather & yes, rather\\ \hline
	9 & yes, rather & yes absolutely & yes, rather\\ \hline
	10 & yes, rather & yes absolutely & yes absolutely\\ \hline
	11 &\cellcolor[HTML]{ff0000} not really & \cellcolor[HTML]{dad500} yes, rather & \cellcolor[HTML]{ff0000} not at all\\ \hline
	12 & yes, rather & yes absolutely & yes absolutely\\ \hline
	13 & yes, rather & yes absolutely & yes, rather\\ \hline
	14 & yes, rather & yes absolutely & yes, rather\\ \hline
	15 & yes absolutely & yes absolutely & yes absolutely\\ \hline
	16 & yes, rather & yes, rather & yes, rather\\ \hline
	17 & yes, rather & yes, rather & yes, rather\\ \hline
	18 & yes, rather & yes, rather & yes, rather\\ \hline
	19 & yes, rather & \cellcolor[HTML]{dad500} yes, rather &\cellcolor[HTML]{fea402}  not really\\ \hline
	20 & \cellcolor[HTML]{fea402} not really & not really & \cellcolor[HTML]{fea402} not really\\ \hline
	21 & \cellcolor[HTML]{fea402} not really & \cellcolor[HTML]{dad500} yes, rather & \cellcolor[HTML]{fea402} not really\\ \hline
	22 & yes, rather & yes, rather & yes, rather\\ \hline
	23 & \cellcolor[HTML]{fea402} not really & \cellcolor[HTML]{dad500} yes, rather & \cellcolor[HTML]{fea402} not really\\ \hline
	24 & \cellcolor[HTML]{fea402} not really & \cellcolor[HTML]{dad500} yes absolutely & \cellcolor[HTML]{fea402} not really\\ \hline
	25 & \cellcolor[HTML]{fea402} not really & not really & \cellcolor[HTML]{fea402} not really\\ \hline
	26 & yes, rather & \cellcolor[HTML]{dad500} yes, rather & \cellcolor[HTML]{fea402} not really\\ \hline
	27 & yes, rather & yes absolutely & yes absolutely\\ \hline
	28 & yes absolutely & yes absolutely & yes absolutely\\ \hline
	29 & yes, rather & yes, rather & yes, rather\\ \hline
	30 & yes absolutely & yes absolutely & yes absolutely\\ \hline
	\end{tabular}
	\label{table4}
\end{table}

\leftskip=0.6cm
\subsubsection{Global analysis}
\label{subsubsecGlobalAnalysis}
\leftskip=0cm

See Table 3. (Question \textbf{Q1}) Twenty-two patients (73.3\%) thought the exercise \textbf{LEARN\_EX} was easy to use after watching the video \textbf{SEQ\_LEARN\_EX}, 
eight patients (26.7\%) felt difficulty to understand the mixed reality environment. (Question \textbf{Q2}) twenty-eight patients (93.3\%) found the communication efficient regarding the 
actions 
to be carried out either in the video sequence \textbf{SEQ\_LEARN\_EX} or in the exercise of learning \textbf{LEARN\_EX}. Only two patients (6.7\%) have experienced technical 
difficulties. (Question \textbf{Q3)} nineteen patients (63.3\%) felt comfortable during the learning exercise, nine patients (30\%) did not feel really comfortable, two patients (6.7\%) not at 
all. For example, one patient showed signs of nervousness because of difficulty to grab the food (too short distance between the hand and the headset).

\leftskip=0.6cm
\subsubsection{Analysis by patient}
\label{subsubsecAnalysis by patient}
\leftskip=0cm

See Table 4. We can notice that among the nine patients (30\%) who were not really comfortable using the learning exercise (orange cell, question \textbf{Q3}), six of them (orange cell, 
question \textbf{Q1}) thought the exercise was not easy to use. One of the two patients (patient id\#2) who expressed not at all comfortable using the exercise (\textbf{Q3}) thought 
the exercise was easy to use. Eleven patients (36.7\%) were not really or at all comfortable and eight of them have answered that the initiation to the technical gestures seemed to them rather 
clear (\textbf{Q2}) and for one of them it was quite clear (green cells, question \textbf{Q3}).

\par Nineteen patients (63.3\%) who were more or somewhat comfortable using the learning exercise (question \textbf{Q3}), eighteen of them found the learning exercise rather easy to use after 
viewing the sequence (question \textbf{Q1}), and that the initiation to the technical gestures seemed to them clear (question \textbf{Q2}), only one patient (patient id\#6, purple cell) thought the 
application was not easy to use and it is the only disconnect between communication and practice.

\par The answers are rather positive concerning the communication (questions \textbf{Q1} and \textbf{Q2}), they are less positive concerning the roll-out of 
\textbf{LEARN\_EX} (question \textbf{Q3}). Additional results concerning the application \textbf{HOLO\_NUTRI}, and not only the communication set up as in this section, were 
also collected \cite{Rositi2020}; the results about this project are based on the NASA-TLX, SUS and on clinical feasibility study CARACO \cite{Barret2019}.

\section{Conclusion}
\label{secConclusion}

Due to its complexity of handling and its novelty, adapted communication was necessary to present the \textbf{HOLO\_NUTRI} hardware and software to thirty patients. The latter found 
the use of this material and the application quite simple thanks to the communication deployed throughout the workshop. Global communication was important and essential. It has been
lightened enough not to overwhelm patients with technological concepts.

\par Despite our experience as computer scientists and communication experts, we did not anticipate as much work was necessary on this part of communication. Finally it represents a 
quarter of the total work done in this experiment; it would have required much less work for a computer application in a conventional setting but the latter would not have allowed a 
similar learning in the gesture of patients and therefore the desired impact of this new workshop.

\par The experience was very much appreciated by both the patients and the teams, it would not have been the case if the communication had not been prepared, thought out and 
carefully presented during the experiment. If we had to deal with a new type of intervention on patients with this type of material in a mixed reality setting, we think that these 
elements (leaflets and videos) are sufficient. More generally, this feedback of communication experience may be of interest to any developer of mixed reality applications.

\section{Acknowledgments}
\label{secAcknowledgments}

\begin{itemize}
\item Grant: PEPS INSIS CNRS " Engineering Sciences for Health to support translational projects ", 2017, AVACM Project (Augmented Visual Assistance during Medical Consultations),
\item Patients of the Bariatric Surgery / Human Nutrition departments, Le Puy-en-Velay, Hospital Center,
\item Jean-Raymond Casimir, Jonathan Etienne, students of the bachelor of Development of Interactive 3D Graphics Applications option, Computer Graphics Department, IUT of Le Puy-en-Velay, for the development of \textbf{HOLO\_NUTRI},
\item Arthur Jacquin, student of  Diploma of Computer Graphics Department, IUT of Le Puy-en-Velay, for the development of \textbf{HOLO\_NUTRI},
\item Marianne Bonneton, Robin Morel, Romain Vergnaud, students of Diploma of Multimedia Department, IUT of Le Puy-en-Velay for the research on visual design of leaflets.

\end{itemize}

\section{Bibliography/References}
\label{secBiblio}
\bibliographystyle{ieeetr} 

\bibliography{avacm}

\begin{thebibliography}{10}

\bibitem{Rositi2020}
H.~Rositi, O.~K. Appadoo, D.~Mestre, S.~Valarier, M.-C. Ombret,
  E.~Gadea-Deschamps, C.~Barret-Grimault, and C.~Lohou, ``Presentation of a
  mixed reality software with a {H}olo{L}ens headset for a nutrition
  workshop.,'' {\em Multimedia Tools and Application}, vol.~80, pp.~1945--1967,
  2021.

\bibitem{Hercberg2004}
S.~Hercberg, P.~Galan, P.~Preziosi, S.~Bertrais, L.~Mennen, D.~Malvy, A.-M.
  Roussel, A.~Favier, and S.~Brian\c{c}on, ``{The SU.VI.MAX Study: A
  Randomized, Placebo-Controlled Trial of the Health Effects of Antioxidant
  Vitamins and Minerals},'' {\em JAMA Internal Medicine}, vol.~164,
  pp.~2335--2342, 11 2004.

\bibitem{Hercberg2004b}
S.~Hercberg, M.~Deheeger, and P.~Preziosi, {\em Portions alimentaires : Manuel
  photos pour l'estimation des quantit\'es}.
\newblock \'Editions Polytechnica, 2002.

\bibitem{hololens2020}
``Microsoft {H}olo{L}ens,'' 2020.

\bibitem{GGV2020}
``{G}aze, {G}esture and {V}oice,'' 2020.

\bibitem{Lohou2019}
C.~Lohou, B.~Miguel, and K.~Azarnoush, ``Preliminary experiment of the
  interactive registration of a trocar for thoracoscopy with {H}olo{L}ens
  headset,'' in {\em Image Analysis and Processing -- ICIAP 2019} (E.~Ricci,
  S.~Rota~Bul{\`o}, C.~Snoek, O.~Lanz, S.~Messelodi, and N.~Sebe, eds.),
  (Cham), pp.~694--703, Springer International Publishing, 2019.

\bibitem{Lohou2019b}
C.~Lohou, M.~Bouiller, and E.~Gadea-Deschamps, ``Mixed reality experiment for
  hemodialysis treatment,'' in {\em Surgetica $9^{th}$ edition}, (Rennes,
  France), June 2019.

\bibitem{Adobe2018}
``Adobe creative suite,'' 2020.

\bibitem{Vuforia2020}
``Vuforia,'' 2020.

\bibitem{Unity2020}
``Unity,'' 2020.

\bibitem{Barret2019}
C.~Barret-Grimault, M.-C. Ombret, O.~K. Appadoo, H.~Rositi, S.~Valarier,
  E.~Privat, I.~Benmabrouk, V.~Haas, V.~Rousset, S.~Verret, C.~Lohou, and
  E.~Gadea, ``Innovons en {ETP} grâce au casque {H}olo{L}ens en chirurgie
  bariatrique,'' in {\em Société d'éducation thérapeutique européenne
  (SETE)}, $7^{e}$ édition, (Toulouse), May 2019.

\end{thebibliography}

\end{document}